\documentclass[12pt]{article}
\usepackage{amsmath}
\usepackage{graphicx}
\usepackage{enumerate}
\usepackage{natbib}
\usepackage{bm}
\usepackage{url} 
\usepackage{booktabs}
\usepackage{amssymb,amsfonts,amsthm}
\usepackage{tikz}
\usetikzlibrary{positioning}
\usetikzlibrary{arrows,arrows.meta}
\RequirePackage[colorlinks,citecolor=blue,urlcolor=blue]{hyperref}

\hypersetup{backref, final=true, pdfpagemode=FullScreen,  colorlinks=true}

\usepackage{amssymb}
\usepackage{longtable}
\usepackage{lineno,hyperref}
\usepackage[section]{placeins}
\usepackage{multirow}
\usepackage{array}
\usepackage[ruled,vlined]{algorithm2e}

\makeatletter
\def\singlespace{\def\baselinestretch{0.7}\@normalsize}
\def\beqn{\begin{eqnarray}}
	\def\eeqn{\end{eqnarray}}
\def\beqns{\begin{eqnarray*}}
	\def\eeqns{\end{eqnarray*}}

\newtheorem{theorem}{Theorem}

\newtheorem{example}{Example}

\newcounter{thm}

\def\bSig\mathbf{\Sigma}

\def\bbeta{\mbox{\boldmath{$\beta$}}}

\def\bmu{\mbox{\boldmath{$\mu$}}}

\def\bSigma{\mbox{\boldmath{$\Sigma$}}}

\def\diag{\mbox{\rm diag}}

\def\bX{\mathbf X}

\def\ba{\mathbf a}

\def\bE{\mathbf E}

\def\bS{\mathbf S}

\def\bV{\mathbf V}

\def\bY{\mathbf Y}
\def\bb{\mathbf b}

\def\bz{\mathbf z}

\def\b0{\mathbf 0}

\def\bbeta{\mbox{\boldmath{$\beta$}}}

\def\bmu{\mbox{\boldmath{$\mu$}}}

\def\bSigma{\mbox{\boldmath{$\Sigma$}}}

\def\diag{\mbox{\rm diag}}

\def\Var{\mbox{\rm Var}}

\def\E{\mbox{\rm E}}

\begin{document}
			\def\spacingset#1{\renewcommand{\baselinestretch}%
		{#1}\small\normalsize} \spacingset{1}
		
	\title{Penalized  Quasi-likelihood for
		High-dimensional Longitudinal Data via Within-cluster Resampling}        
	\author{\hspace{0.7cm}Yue Ma$^1$, Haofeng Wang$^2$ and Xuejun Jiang$^{1,}$\footnote{Corresponding author}     \\ 
		\vspace{0.1cm}
		\hspace{ 1cm}\textsl{ $^1$Department of Statistics and Data Science, Southern University  
		} \\
		\hspace{ 0.1cm} \textsl{of Science  and Technology, China}\\
		\vspace{0.1cm}
		\hspace{ 0.2cm}\textsl{ $^2$Department of Mathematics, Hong Kong Baptist University,
		} \\
		\hspace{ 0.2cm} \textsl{Hong Kong}
	}

	\date{}      
	\maketitle
	\noindent{\bf Abstract:}
The generalized estimating equation (GEE) method is a popular tool for longitudinal data analysis. However, GEE produces biased estimates when the outcome of interest is associated with cluster size, a phenomenon known as informative cluster size (ICS). In this study, we address this issue by formulating the impact of ICS and proposing an integrated approach to mitigate its effects. Our method combines the concept of within-cluster resampling with a penalized quasi-likelihood framework applied to each resampled dataset, ensuring consistency in model selection and estimation. To aggregate the estimators from the resampled datasets, we introduce a penalized mean regression technique, resulting in a final estimator that improves true positive discovery rates while reducing false positives. Simulation studies and an application to yeast cell-cycle gene expression data demonstrate the excellent performance of the proposed penalized quasi-likelihood method via within-cluster resampling.

	~\\
	
	
	\newpage
	\spacingset{1.9}
\section{Introduction}\label{sec1}

Longitudinal data, characterized by repeated and correlated observations across multiple covariates, frequently arise in diverse fields such as medical research, epidemiology, and the social sciences. \cite{FDVM2008} highlighted examples from these disciplines and discusses statistical approaches for analyzing such data. Conventional methods designed for independent data often fail to account for the dependence among longitudinal observations, resulting in inefficient estimates and invalid statistical inferences.  Thus, special techniques to address correlated measurements must be developed urgently.      

The generalized estimating equation (GEE) method \citep{Liang} is one of the most widely used techniques for longitudinal data analysis. It accounts for the dependence among observations through a working correlation matrix and can produce robust solutions regardless of whether the within-cluster correlation structure is correctly specified. However, this robustness is compromised when observations within a cluster are not completely missing at random.  For instance, in a periodontal study \citep{GWS1998}, patients with poor dental health tend to have fewer teeth than those in good oral condition while also facing a higher risk of dental diseases. Let $X$ denote  the covariates, $Y$ the disease status of each tooth (response), and $M$ the cluster size (number of teeth). Here, the response $Y$ is influenced by the cluster size $M$, a phenomenon known as informative cluster size (ICS) \citep{Hoffman2001}, which can be expressed as $E(Y|X, M)\not=E(Y|X)$. Another example arises in toxicology experiments where pregnant dams are randomly exposed to toxicants. Sensitive dams may produce litters with a higher rate of birth defects and experience more fetal resorptions, resulting in smaller litter sizes.

To address these challenges, \cite{Hoffman2001} proposed the within-cluster resampling (WCR) approach for marginal analysis of longitudinal data. This method involves sampling one observation from each cluster with replacement and then performing statistical inference using techniques developed for independent data. WCR offers two notable advantages over the GEE method. First, it implicitly accounts for within-cluster correlation, thereby avoiding biases that arise from misspecified working correlation structures. Such biases are a known limitation of GEE, which can yield inaccurate estimates when the assumed correlation matrix is incorrect \citep{ML1996}, or when the true dependence structure is too complex to model practically. Second, WCR remains valid in the presence of ICS, in contrast to GEE, which assigns greater weight to observations from larger clusters. This weighting introduces inefficiencies when cluster sizes carry information about the response. By contrast, WCR eliminates the influence of ICS by conducting analyses on independent resampled data from each cluster, achieving consistent estimation regardless of cluster size informativeness.

Recent research has increasingly focused on high-dimensional longitudinal data analysis, particularly addressing challenges in variable selection and parameter estimation. \cite{Wang2009} introduced a Bayesian information criterion (BIC) type model selection criterion based on the quadratic inference function,
while \cite{Wang2011} developed an asymptotic theory for binary outcomes when the number of predictors grows with the number of clusters. The penalized GEE method \citep{Wang2012} utilizes the smoothly clipped absolute deviation penalty (SCAD) penalty \citep{Fan2001} to perform variable selection and estimation simultaneously, provided the first two marginal moments and a working correlation structure are prespecified.  Additionally, \cite{FNL20} proposed a quadratic decorrelated inference function approach for inferences in high-dimensional longitudinal data, and Xia and Shojaie (2023) constructed a one-step de-biased estimator using projected estimating equations for GEE, enabling the testing of linear combinations of high-dimensional regression coefficients in longitudinal settings.

However, these GEE-based approaches rely on estimating within-cluster correlation, which leads to inconsistent performance when ICS is present.
To address this, estimation methods under the WCR framework have been developed. For example,
\cite{CL2008} proposed the modified WCR  approach to enhance estimation efficiency, though it requires clusters to have a minimum size greater than one and imposes additional restrictions on the correlation structure. \cite{SC2018} introduced a resampling cluster information criterion for model selection in semiparametric marginal mean regression with finite dimensions. However, the above WCR methods rely on averaging multiple resampled-based estimators. This averaging approach lacks theoretical guarantees for model selection in high-dimensional regression, potentially leading to overfitting \citep{Wel23}.

In this study, we propose a model selection procedure for high-dimensional longitudinal data in the presence of ICS. The dimension of covariates $p_n$ is allowed to grow exponentially with the number of clusters $n$. Building on the stability that WCR ensures in finite dimensions when ICS is present, we aim to explore its model selection and estimation consistency in scenarios where the number of covariates exceeds the total number of clusters. The proposed method comprises three main steps as follows: First, following the WCR framework \citep{Hoffman2001}, we randomly sample an observation from each cluster with replacement. Second, using the sampled $n$ independent observations, we perform  variable selection and estimation by maximizing the penalized quasi-likelihood to obtain a consistent estimator. This process is repeated $K$ times as part of a Monte Carlo procedure. Finally, to mitigate the risk of severe overfitting from averaging across the $K$ estimates, we apply penalized mean regression component-wise to these estimates, producing a sparse final model. This methodology is referred to as the penalized quasi-likelihood via WCR, abbreviated as PQL$_{\rm WCR}$, because it combines penalized quasi-likelihood with the WCR approach.

Compared with existing variable selection methods for longitudinal data, the PQL$_{\rm WCR}$ method offers three key advantages. (i) Implicitly handling intracluster correlation: PQL$_{\rm WCR}$ addresses intracluster correlation indirectly, thereby avoiding the invalid estimates that can arise from misspecified within-cluster correlation structures, particularly in high-dimensional settings. (ii) Robustness to ICS: the method remains robust even when dependencies exist between the responses and cluster sizes, effectively managing the challenges posed by ICS. (iii) Sparse solutions via penalized mean regression: by leveraging penalized mean regression during the aggregation process, PQL$_{\rm WCR}$  achieves sparse model solutions, greatly reducing the risk of overfitting.

We organize this paper as follows: Section \ref{sec2} introduces the model assumptions, discusses the biases of the GEE estimator, and provides a detailed explanation of the PQL$_{\rm WCR}$ method. In Section \ref{the}, we outline the regularity conditions and establish the theoretical properties of the proposed method. Section \ref{simu} presents simulation studies and comparative analyses with two competing approaches. In Section \ref{real}, we demonstrate the application of our method to a yeast cell-cycle gene expression dataset. In Section \ref{discu}, we provide a brief discussion. The proofs of the theorems are delegated into the Appendix.

Throughout this article, we use the following notations. For any constants $a$, $[a]$ denotes its integer part. For vectors $\ba=(a_{1},\ldots,a_{m})^T\in \mathbb{R}^m$ and $\bb=(b_{1},\ldots,b_{m})^T\in \mathbb{R}^m$, we define $|\ba|=(|a_1|,\ldots,|a_m|)^T$, $\Vert\ba\Vert_{1}=\sum_{i=1}^{m}|a_{i}|$, $\Vert \ba\Vert_{2}=\sqrt{\sum_{i=1}^{m}a_{i}^2}$, $|\ba|_{\infty}=\max_{1\leq i\leq m}|a_i|$, and $\ba\circ\bb=(a_1b_1,\ldots,a_m$
$b_m)^T$. For any subset $\mathcal{M}$ of the row index set  $\{1,\ldots,m\}$ of $\ba$,  $|\mathcal{M}|$ denotes the cardinality of $\mathcal{M}$, and $\mathbf{a}_{\mathcal{M}}$ represents the subvector of $\ba$ corresponding to the indices in $\mathcal{M}$. The indicator function is denoted by $I(\cdot)$. Let $\bE$ be a  $m\times m$ matrix  and $\bE^{-1}$ its inverse. For subsets $\mathcal{M}_1,\mathcal{M}_2\subseteq \{1,\ldots,m\}$, $\bE_{\mathcal{M}_1}$ denotes the submatrix of $\bE$ formed by column indices in $\mathcal{M}_{1}$, and $\bE_{\mathcal{M}_{1},\mathcal{M}_{2}}$ represents the submatrix whose row and column  are indexed by $\mathcal{M}_{1}$ and $\mathcal{M}_{2}$, respectively. The maximum and minimum eigenvalues of a matrix are denoted by $\lambda_{\max}(\cdot)$ and $\lambda_{\min}(\cdot)$, respectively, and the spectral norm is given by $\|\bE\|_{2}=\sqrt{\lambda_{\max}(\bE^T\bE)}$.

\section{Methodology}\label{sec2}
\subsection{Problem formulation}
\hspace{3mm} Let $Y_{i j}$ represents the response variable of the $j$th observation in the $i$th cluster, with a $p_n$-dimensional covariate vector  $\mathbf{X}_{ij}=(X_{ij,1},\dots,X_{ij,p_n})^T$, where $i=1,\dots,n$, $j=1,\dots,M_i$, and $M_i$ refers to cluster size. We consider the first two marginal moments of $Y_{i j}$ as follows:
\begin{equation}\label{MT1}
	\E(Y_{ij}\mid \bX_{ij})=\mu(\bX_{ij}^T\bbeta^*) \ \ \text{and} \ \
	\Var(Y_{ij}\mid \bX_{ij})=\phi g(\mu(\bX_{ij}^T\bbeta^*)),
\end{equation}
where $\bbeta^*$ is an unknown $\mathbb{R}^{p_n}$ coefficient vector,
$\mu$ is a known link function, $g(\cdot)$ is the variance function,  and
$\phi$ is the dispersion parameter. For simplicity, we assume $\phi=1$ in the remainder of this article.
Observations within each cluster are commonly assumed to be correlated,  i.e. $|{\rm Corr}(Y_{ij}, Y_{ik})|>0$, but observations are independent across clusters. Here, we allow the dimensionality $p_n$  to be larger than the number of culsters $n$, i.e, $\log(p_n)=n^{\alpha}$ for $0<\alpha<1$, and the true population parameter $\bbeta^*$ is sparse with only several nonzero components. 
Define $\bmu_i(\bbeta)=(\mu(\bX_{i1}^T\bbeta),\ldots,$
$\mu(\bX_{iM_i}^T\bbeta))^T$. Let   $\bY_{i}= (Y_{i1},Y_{i2},\dots,Y_{iM_i})^T$ and $\bX_i=(\bX_{i1},\bX_{i2},\dots,\bX_{iM_i})^T$.

To obtain the optimal estimation, \cite{Liang} proposed the GEE:
\begin{equation}\label{MT2}
	n^{-1}\sum_{i=1}^n \bigl\{\bX_i^T\bmu_i'(\bbeta) \bV_i^{-1}
	(\bY_{i}-\bmu_i(\bbeta))\bigr\}=\b0,
\end{equation}
where $\bV_i$ is the covariance matrix of $\bY_i$,  $\bmu_i'(\bbeta)=\diag(\mu'(\bX_{i1}^T\bbeta),\ldots,$$\mu'(\bX_{iM_i}^T\bbeta))$ 
with $\mu'(\cdot)$ being the derivative of $\mu(\cdot)$. A naive choice of the covariance matrix is the independent structure, i.e. $\bV_i=\diag(g(\mu(\bX_{i1}^T\bbeta)),\ldots,g(\mu(\bX_{iM_i}^T\bbeta)))$.  However, the validness of GEE is built upon the assumption that $M_i$ is independent of $\{Y_{ij},\bX_{ij},i=1,\ldots,n,j=1,\ldots,M_i\}$. This assumption is violated when the cluster size is informative \citep{Hoffman2001}, i.e.  $E(Y_{ij}\mid \bX_{ij}, M_{i})\neq E(Y_{ij}\mid \bX_{ij})$. With the presence of ICS, it follows that
\begin{equation}\label{sec2.5}
	E\{\bX_i^T\bmu_i'(\bbeta^*) \bV_i^{-1}
	(\bY_{i}-\bmu_i(\bbeta^*))\}
	=E\bigl\{M_i E\bigl(f(\bX_{i1})\epsilon_{i1}\mid M_i\bigr)\bigr\}\neq \b0,
\end{equation} 
by the law of iterated expectation, where $f(\bX_{i1})=\bX_{i1}\mu'(\bX_{i1}^T\bbeta^*)/g(\mu(\bX_{i1}^T\bbeta^*))$ and
$\epsilon_{i1}=Y_{i1}-\mu(\bX_{i1}^T\bbeta^*)$. 
From  \eqref{sec2.5}, the GEE leads to a biased estimation. This motivates us to propose a novel estimation method for longitudinal data when the cluster size is informative.

\subsection{Penalized WCR quasi-likelihood estimation }
For model (\ref{MT1}), the quasi-likelihood function of $\{Y_{ij},\bX_{ij},i=1,\ldots,n,j=1,\ldots,M_i\}$ is 
\begin{equation}\label{sec2.6}
	Q_{n}(\bbeta)=n^{-1}\sum_{i=1}^n\sum_{j=1}^{M_i}Q(Y_{ij},\bX_{ij}^T\bbeta),
\end{equation}
where $Q(Y_{ij},\bX_{ij}^T\bbeta)=\int_{\mu(\bX_{ij}^T\bbeta)}^{Y_{ij}}
(s-Y_{ij})g^{-1}(s) ds$.
By taking derivative, we obtain
$$\partial Q_{n}(\bbeta)/\partial\bbeta=n^{-1}\sum_{i=1}^n \bigl\{\bX_i^T\bmu_i'(\bbeta) \bV_i^{-1}
(\bY_{i}-\bmu_i(\bbeta))\bigr\},$$
where $\bV_i=\diag(g(\mu(\bX_{i1}^T\bbeta)),\ldots,g(\mu(\bX_{iM_i}^T\bbeta)))$ is the assumed working independence matrix. The GEE often prespecifies different working correlation structures in $\bV_i$ to increase efficiency, such as exchangeable or autoregressive structures. 

If we let $\partial Q_{n}(\bbeta)/\partial\bbeta=\b0$, it is exactly the GEE equation (\ref{MT2}). However, this does not happen when the ICS is presented as demonstrated by  (\ref{sec2.5}). The effect of ICS originates from the randomness of $\sum_{j=1}^{M_i}Q(Y_{ij},\bX_{ij}^T\bbeta)$ in (\ref{sec2.6}). To eliminate the effect of ICS, we adopt the WCR idea proposed by \cite{Hoffman2001}. 
Let $Z_i$ be a random variable independent of $\{Y_{ij},\bX_{ij},i=1,\ldots,n,j=1,\ldots,M_i\}$ and $z_i$ is the sample of $Z_i$, which takes values in $\{1,\ldots,M_i\}$
with the probability $P(z_i=j)=1/M_i$, $j=1,\ldots,M_i$. WCR randomly draws an observation from each cluster with replacement. Thus, the quasi-likelihood function via WCR is 
\begin{equation}\label{sec27}
	\tilde{Q}_{n}(\bbeta;\bz)=n^{-1}\sum_{i=1}^n\sum_{j=1}^{M_i}I(j=z_i)Q(Y_{ij},\bX_{ij}^T\bbeta),
\end{equation}
where $I(\cdot)$ is the indicator function and $\bz=(z_1,\ldots,z_n)$. With straightforward calculations and the use of the law of iterative expectation, we can obtain
$$E\{\partial \tilde{Q}_{n}(\bbeta^*;\bz)/\partial\bbeta\}=
E\bigl(f(\bX_{iz_i})\epsilon_{iz_i}\bigr)=\b0.
$$
This guarantees the resulting maximum quasi-likelihood estimation is consistent. Benefited from this property, we develop variable selection and simultaneous estimation for model (\ref{MT1}).
The penalized quasi-likelihood function via WCR  is defined as
\begin{align}\label{plf}
	\ell(\bbeta;\bz)=\tilde{Q}_{n}(\bbeta;\bz)
	-\sum_{d=1}^{p_n}p_{\lambda_{n}}(|\beta_{d}|),
\end{align}
where $p_{\lambda_{n}}(\cdot)$ is a penality function indexed by a regularization parameter $\lambda_n$. We generate the $K$ independent observations $\{\bz^{(1)},\ldots,\bz^{(K)}\}$ of $\bz$.  
Let $\tilde{\bbeta}^k$ be the maximizer of $\ell(\bbeta,\bz^{(k)})$, i.e.
\begin{equation}\label{plfe}
	\tilde{\bbeta}^{(k)}=
	\arg\max\limits_{\footnotesize\bbeta}\ell(\bm{\beta};\bz^{(k)}),
\end{equation}
for $k=1,\ldots,K$.
How to integrate the $K$ estimators  is the next issue we need to focus on. A naive approach is to average the $K$ estimators, as proposed in \cite{Hoffman2001}. However, this trivial averaging approach may cause many covariates to have nonzero estimated coefficients, resulting in  model overfitting. Alternatively, we propose to  aggregate these $K$ estimators by regularization. Therefore, we consider the following penalized  least squares estimation:
\begin{equation}\label{AFFN}
	\hat\bbeta=\arg\min\limits_{\footnotesize \bbeta}
	\{K^{-1}\sum_{k=1}^K\|\tilde{\bbeta}^{(k)}-\bbeta\|_{2}^2+\sum_{d=1}^{p_n}\lambda_{n}^{(d)}|\beta_{d}|\}.
\end{equation}
Here, we summarize our penalized quasi-likelihood (PQL) via WCR (PQL$_{\rm WCR}$) method in Algorithm 1.

\begin{algorithm}
	\caption{PQL$_{\rm WCR}$ algorithm}
	\label{PCRL algorithm}
	\KwIn{$\{(\bY_i,\bX_i)\}_{i=1}^n$, $\lambda_{n}$ and $\{\lambda_{n}^{(1)},\ldots,\lambda_{n}^{(p_n)}\}$}
	\KwResult{$\hat{\mathbf{S}}$ and $\hat{\bbeta}$}
	\textbf{Step 1: Penalized  quasi-likelihood estimation via WCR}\\
	\For {$k=1,\dots,K$,}{
		Randomly draw an observation $\bz^{(k)}$\;
		Compute the estimator $\tilde{\bbeta}^{(k)}$ by solving\
		$\tilde{\bbeta}^{(k)}={\arg\max}_{\footnotesize \bbeta}\bigl\{\tilde{Q}_{n}(\bbeta;\bz)
		-\sum_{d=1}^{p_n}p_{\lambda_{n}}(|\beta_{d}|)\}$\;
	}		
	\textbf{Step 2: Aggregation}\\
	On the basis of $\{\tilde{\bbeta}^{(k)}\}^K_{k=1}$, calculate the aggregate estimates $\hat{\bbeta}$ as
	$\hat{\bbeta}={\arg\min}_{\footnotesize \bbeta}
	\{K^{-1}\sum_{k=1}^K\|\tilde{\bbeta}^{(k)}-\bbeta\|_{2}^2+\sum_{d=1}^{p_n}\lambda_{n}^{(d)}|\beta_{d}|\}$\;
	
	Then let $\hat{\mathbf{S}}=\{d:|\hat{\beta}_d|>0\}$.\\
\end{algorithm}

\section{Theoretical properties}\label{the}
In this section, we establish the consistency in model selection and parameter estimation.

For convenience, we first introduce some notations
Let $\bS$ be the index set corresponding to the nonzero components in $\bbeta^*$. Define $\varepsilon_{ij}=Y_{ij}-\mu(\bX_{ij}^T\bbeta)\mu'(\bX_{ij}^T\bbeta)g^{-1}(\mu(\bX_{ij}^T\bbeta))$, $\mathcal{N}_0=\{\bbeta:\, \|\bbeta-\bbeta^*\|_{2}\leq \sqrt{s_n/n},\bbeta_{\bS^c}=\b0\}$, and $\mathcal{P}=\{\bz=(z_1,\ldots,z_n):\,z_{i}\in \{1,2\ldots,M_i\},i=1,\ldots,n\}$.
Denote $\bSigma_{1}(\bbeta;\bz)=\partial^2 \tilde{Q}_{n}(\bbeta;\bz)/\partial\bbeta_{\bS}\partial\bbeta_{\bS}^T$ and
$\bSigma_{21}(\bbeta;\bz)=\tilde{Q}_{n}(\bbeta;\bz)/\partial\bbeta_{\bS^c}\partial\bbeta_{\bS}^T$. Let $s_n=|\bS|$ and $d_n=\min_{d\in\bS}|\beta_d^*|/2$. 

The following regularity conditions are required for establishing asymptotic results.
\begin{itemize}
	\item[(A1)]
	Let $\rho_{\lambda_{n}}(t)=\lambda_n^{-1}p_{\lambda_n}(t)$  for $\lambda_n>0$. Assume that $\rho_{\lambda_{n}}(t)$ is increasing and concave in $t\in[0,\infty)$ and has a continuous derivative $\rho_{\lambda_{n}}'(t)$ with $\rho_{\lambda_{n}}'(0+)>0$.
	In addition,
	$\rho_{\lambda_{n}}'(0+)$ is independent of $\lambda_n$,
	and for simplicity, we write it as $\rho'(0+)$.
	
	\item[(A2)] 
	(i) $\max_{1\leq i\leq n}\max_{1\leq j\leq M_i} E\big\{\exp(|\varepsilon_{ij}|/c_{1})-1-|\varepsilon_{ij}|/c_{1}\mid\bX_{ij}\}c_{1}^2\leq \frac{c_{2}}{2}$ for some positive constants $c_{1}$ and $c_2$;\\
	(ii)
	$\|\ba^T\bX_{ij}\|_{\psi_{2}}\leq c_{6}\|\ba\|_{2}$ for any $\ba\in R^{p_n}$,
	where $\|\ba^T\bX_{ij}\|_{\psi_{2}}=\inf_{t>0}\{E\exp(t^{-2}|\ba^T\bX_{ij}|^2)\leq 2\}$. 
	\item[(A3)] Assume that for any $\bz\in\mathcal{P}$, there exist constants $c_3,c_4>0$ such that
	$\min_{\bbeta\in\mathcal{N}_0}\lambda_{\min}(\bSigma_{1}(\bbeta;\bz))\geq c_3$ and $\max_{\bbeta\in\mathcal{N}_0}\sup_{\ba\in R^{s_n},\|\ba\|_{2}\leq 1}\|\bSigma_{21}(\bbeta;\bz)\ba\|_{\infty}\leq c_4$ holds with probability going to one.
	
	\item[(A4)] Assume that $\lambda_{n}=o(d_n)$, $\max\{\sqrt{s_n/n},\sqrt{\log(p_n)/n}\}=o(\lambda_{n})$ and $p_{\lambda_{n}}'(d_n)=O(n^{-1/2})$.
	
	\item[(A5)] Define the supremum of the local concavity of $\rho_{\lambda_{n}}$ as
	$$\kappa(\rho_{\lambda_n})=\sup_{{\bb}\in\mathcal{N}_0}\lim_{a\rightarrow 0+}\max_{ k\in \bS}
	\sup_{t_{1}<t_{2}\in(|b_k|-a,|b_k|+a)}-\frac{\rho_{\lambda_n}'(t_{2})-\rho_{\lambda_n}'(t_{1})}{t_{2}-t_{1}}$$
	and suppose $\lambda_n\kappa(\rho_{\lambda_n})=o(1)$.
\end{itemize} 

Condition (A1) is satisfied by commonly used folded concave penalty functions, such as LASSO \citep{Tibshirani1996}, SCAD \citep{Fan2001}, and MCP \citep{Zhang2010} (with $a\geq 1$). \cite{Fan2001} states that an ideal estimator should be equipped with three desirable properties: unbiasedness, sparsity, and continuity. However, the LASSO penalty fails to reach an unbiased estimator, the $L_q$ penalty with $q>1$ does not produce a sparse solution, and the $L_q$ penalty with $0\leq q<1$ does not satisfy the continuous condition. 
In our study, SCAD penalty is chosen  to conduct variable selection, whose derivative is defined as
\begin{equation*}
	p^\prime_{\lambda_n}(t)=\lambda_n\left\{I(t\leq\lambda_n)+\frac{(a\lambda_n-t)_+}{(a-1)\lambda_n}I(t>\lambda_n)\right\}, 
\end{equation*}
for some $a>2$ and $t\geq 0$. This penalty satisfies both Condition (A1) and the three desirable properties within penalized likelihood framework. 

Condition (A2) restricts the moment of errors and the sub-Gaussian distribution of the predictors satisfies this condition, which is commonly used for high dimensional penalized regression \citep{FL11, Sel18,Sel19}. Condition (A2) (i) holds for the exponential generalized linear model family.

In Condition (A3), the constraint 	$\min_{\bbeta\in\mathcal{N}_0}\lambda_{\min}(\bSigma_{1}(\bbeta,\bz))\geq c_1$ and $\max_{\bbeta\in\mathcal{N}_0}\sup_{\ba\in R^{s_n},\|\ba\|_{2}\leq 1}$
$\|\bSigma_{21}(\bbeta,\bz)\ba\|_{\infty}\leq c_4$ is the same as  $(26)$ and $(27)$ in \cite{FL11}, which has been proven for exponential generalized linear model family in \cite{Sel19} if the covariate vector follows from a sub-Gaussian distribution. 
Furthermore, $\|\bSigma_{21}(\bbeta,\bz)\ba\|_{\infty}\leq c_4$ is
the irrepresentable condition \citep{Zhao2006}, which is commonly assumed for penalized regression.

Condition (A4) restricts the strength of signals, the number of nonzero covariates, the dimension of feature space, and the tuning parameter, which is commonly used in \cite{FL11} and \cite{Sel19}. From Condition (A4), we allow $\log(p_n)=n^{\alpha}$, $0<\alpha<1$, which is a common  condition for  the ultra-high dimensional feature space in the literature.

Condition (A5) is assumed by \cite{FL11} to ensure the second-order condition.

The following theorem establishes the oracle property of penalized  quasi-likelihood estimator $\tilde{\bbeta}^{(k)}$ via WCR. 

\begin{theorem}\label{Th2}
	{\rm
		Assume Conditions (A1)--(A5) hold. Then we have that 
		$\|\tilde{\bbeta}^{(k)}-\bbeta^*\|_{2}=O_{p}(\sqrt{s_n/n})$
		and $P(\hat\bS^{(k)}=\bS^*)\to 1$ as $n\to \infty$,  
		where $\hat\bS^{(k)}=\{d:
		\,\tilde{\beta}_{d}^{(k)} \neq 0, d=1,\ldots,p_n\}$. 
	}
\end{theorem}
Theorem \ref{Th2} establishes model selection sparsity and the estimation consistency of $\tilde{\bbeta}^{(k)}$. It also holds for the ultra-high dimensional feature space whether the cluster size is informative or not.

In the following, we establish the oracle property of penalized  quasi-likelihood estimator $\hat{\bbeta}$ via WCR.
\begin{theorem}\label{Th3}
	{\rm
		Assume Conditions in Theorem \ref{Th2} hold. If 
		$\lambda_{n}^{(d)}$ satisfy $\lambda_{n}^{(d)}=o(d_n)$ and
		$\sqrt{s_n/n}=o(\lambda_{n}^{(d)})$, then
		we have that 
		$\|\hat{\bbeta}-\bbeta^*\|_{2}=O_{p}(\sqrt{s_n/n})$
		and $P(\hat\bS=\bS^*)\to 1$ as $n\to \infty$,  
		where $\hat\bS=\{d:
		\,\hat{\beta}_d\neq 0, d=1,\ldots,p_n\}$. 
	}
\end{theorem}
Compared with the penalized GEE proposed by \cite{Wang2012}, the PQL$_{\rm WCR}$ algorithm has some significant advantages. On the one hand,  PQL$_{\rm WCR}$  improves the accuracy of model selection, especially when ICS is present. As described in \cite{Hoffman2001}, WCR assigns equal weights on observations from varying sizes clusters to ensure robustness when cluster size is related to model parameters.  
Through numerous marginal analyses by running PQL on resampled datasets, we repeatedly take advantage of sample information and individually achieve consistent screening results. Afterward, we aggregate the $K$ candidate models by penalized least squares, that is,  we conduct penalized mean regression on estimates from the $K$ individual marginal analysis component by component. 
When an important covariate is dropped out by one marginal analysis, fortunately, its aggregated penalized estimate will not be extensively influenced. In addition, several misidentifications of a redundant covariate will not lead its aggregated estimation far from zero.   
That is, PQL$_{\rm WCR}$  can simultaneously increase the true positive discovery rate and decrease the false positive discovery rate. By contrast, the penalized GEE (PGEE) performs only a single fitting on the data, giving each variable only one chance to be selected. This lack of iterative refinement makes it less robust because a single fitting error can jeopardize the final consistency of the results. In addition, the PGEE tends to select plenty of unimportant predictors  in ultra high-dimensional model settings, leading to model overfitting.

On the other hand, the PQL$_{\rm WCR}$ method addresses the within-cluster correlation implicitly.
The within-cluster resampling is a simple but computationally intensive approach. he unknown and perhaps complex intracluster correlation structure do not need to be estimated. However, the PGEE relies heavily on estimating within-cluster correlation, whose estimate is usually invalid for non-linear cases. 

\section{Simulation studies}\label{simu}
In this section, we conduct several numerical experiments to assess the performance of our  PQL$_{\rm WCR}$ method in model selection for longitudinal data. 
We repeat the sampling procedure with replacement for $K=500$ times.
The tuning parameter in the SCAD penalty function is determined by minimizing BIC \citep{Schwarz1978}.

We consider two competing approaches here. 
The first is the PGEE proposed by \cite{Wang2012}, using R package \texttt{PGEE}. Three commonly used working correlation structures are considered, including independence, exchangeable (equally correlated), and autocorrelation (AR)-1. To facilitate the comparison of various
methods under different correlation scenarios, we use the suffixes ``.indep," ``.exch" and ``.ar1" to represent the three correlation structures mentioned above, respectively. As suggested by \cite{Wang2012}, we run the PGEE in 30 iterations and use a fourfold cross-validation to select the tuning parameter in the SCAD penalty function. Following \cite{Wang2012}, we recognize a coefficient as zero if its estimated magnitude is smaller than the threshold value $10^{-3}$. 
In addition, we consider neglecting the within-cluster correlation and conduct penalized maximum likelihood via the $L_1$ penalty, called naive lasso. 	
Investigations include correlated continuous and binary responses.
The PGEE spends about 4 hours on fitting a single generated data set even if $p_n=50$. However, our  PQL$_{\rm WCR}$ method only needs 5 minutes to finish a task. Taking time into consideration, we omit the performance of the penalized GEE in $p_n=500$.

We generate 100 longitudinal datasets for each setup. 
To evaluate the screening performance, we calculate the following criteria: (1) True positive (TP): the average number of true variables that are correctly identified; (2) False positive (FP): the average number of unimportant variables that are selected by mistake; (3) Coverage rate (CR): the probability that the selected model covers the true model;  (4) Mean square error (MSE): calculated as $\sum_{q=1}^{100}\Vert\widehat{\bm{\beta}}^q-\bm{\beta}\Vert^2/100$, where $\widehat{\bm{\beta}}^q$ is the estimate obtained on the $q$th generated dataset. All the algorithms are run on a computer with Inter(R) Xeon(R) Gold 6142 CPU and 256 GB RAM.

\begin{example}\label{ex1}
	{\rm In this example, we consider the underlying model as
		\begin{equation*}
			Y_{ij}=U_{ij}\bX^T_{ij}\bm{\beta}+\epsilon_{ij},    
		\end{equation*}
		for correlated normal responses with ICS, where $i=1,\dots,200$ and $j=1,\dots,M_i$. 
		The random cluster size $M_i$ takes value in the set $\{2,4,15\}$, with the probability distribution as $P(M_i=2)=9/16$, $P(M_i=4)=3/8$, and $P(M_i=15)=1/16$. 
		We define $U_{ij}=1(M_i\leq 4)$.
		The $p_n$-dimensional vector of covariates $\bX_{ij}=(x_{ij,1},\dots,x_{ij,p_n})^T$ are independently generated from a multivariate normal distribution, which has mean 0 and an autoregressive covariance matrix with marginal variance 1 and autocorrelation $0.4$. 			
		The coefficient vector $\bm{\beta}=(2,-1,1.5,-2,0,0,\dots,0)^T$ presents a homogeneous effect. 			
		The random errors $(\epsilon_{i1},\dots,\epsilon_{iM_i})^T$ obeys a multivariate normal distribution with marginal mean 0, marginal variance 1, and an exchangeable correlation matrix with parameter $\rho=0.5\ {\rm or}\ 0.8$.}
\end{example} 

Table \ref{continuous1} summarizes the model selection results and estimation accuracy of three competing approaches for Example \ref{ex1}. All methods can correctly identify TPs but only the  PQL$_{\rm WCR}$ method almost selects the true model. For $p_n=50$, the PGEE selects several redundant variables when observations are assumed to be independent, and identifies a dozen FPs under exchangeable and autoregressive correlation structures. Whereas, naive lasso selects a fairly moderate size model. Furthermore, our  PQL$_{\rm WCR}$ achieves the smallest MSE while maintaining robustness against increases in model dimension and intracluster correlation. In contrast, the estimation errors of the PGEE highlight its inconsistency in the presence of intracluster correlation, with errors doubling as within-cluster correlation increases. In addition,  naive lasso shows biased estimates and performs sensitively to the dimension of covariates, which leads to a larger MSE in estimation.

\begin{table}
	\centering
	\caption{Model selection results for Example \ref{ex1} with ICS.}
	\label{continuous1}
	\renewcommand\arraystretch{0.8}
		\vspace{0.2cm}
	\begin{tabular}{ccccccc}
		\toprule 
		$p_n$&$\rho$	& Approach& TP &FP & CR &MSE \\
		\midrule 
		
		50&$0.5$&PQL$_{\rm WCR}$ & 4.00(0.00) & 0.85(1.08) & 1.00(0.00)  & 0.033(0.022)  \\
		~&~&PGEE.indep& 4.00(0.00) & 7.13(3.92)  & 1.00(0.00) & 0.395(0.117) \\
		~&~&PGEE.exch& 4.00(0.00) & 14.79(9.35)  & 1.00(0.00) & 0.596(0.287)  \\
		~&~&PGEE.ar1& 4.00(0.00) & 16.29(8.40) & 1.00(0.00) & 0.513(0.143)  \\
		~&~&naive lasso& 4.00(0.00) & 2.55(1.95)  & 1.00(0.00) & 0.979(0.215)  \\
		
		~&$0.8$&PQL$_{\rm WCR}$ & 4.00(0.00) & 0.70(0.88) & 1.00(0.00)  & 0.032(0.021) \\
		~&~&PGEE.indep& 4.00(0.00) & 6.69(3.18) & 1.00(0.00) & 0.406(0.109)  \\
		~&~&PGEE.exch&4.00(0.00) & 19.44(13.51) & 1.00(0.00) & 1.099(1.139)  \\
		~&~&PGEE.ar1&4.00(0.00) & 21.31(14.05) & 1.00(0.00) & 1.166(1.066)  \\
		~&~&naive lasso& 4.00(0.00) & 2.69(2.41)  & 1.00(0.00) & 0.978(0.225)  \\
		\hline
		500&$0.5$&PQL$_{\rm WCR}$ & 4.00(0.00) & 0.19(0.44) & 1.00(0.00)  & 0.032(0.018) \\
		~&~&naive lasso& 4.00(0.00) & 2.93(2.01)  & 1.00(0.00) & 1.430(0.278)  \\
		
		~&$0.8$&PQL$_{\rm WCR}$ &4.00(0.00) & 0.27(0.57) & 1.00(0.00)  & 0.032(0.019) \\
		~&~&naive lasso& 4.00(0.00) & 3.09(2.19)  & 1.00(0.00) & 1.426(0.295)  \\
		\bottomrule
		
	\end{tabular}
\end{table}

\begin{example}\label{ex2}
	{\rm We consider correlated binary responses with marginal mean, satisfying
		\begin{equation*}
			E(Y_{ij}\mid \bX_{ij},M_i)=U_{ij}\exp(\bX^T_{ij}\bm{\beta})/\{1+\exp(\bX^T_{ij}\bm{\beta}
			+\log(15/16))\},
		\end{equation*} 
		where $i=1,\dots,200$ and $j=1,\dots,M_i$. To account for ICS, we let the random cluster size $M_i$ follow the probability distribution as $P(M_i=4)=9/16$, $P(M_i=6)=3/8$, and $P(M_i=10)=1/16$.
		The components of covariates vector $\bX_{ij}=(x_{ij,1},\dots,x_{ij,p_n})^T$ are jointly normal distributed with mean 0 and an autoregressive covariance matrix. The covariance matrix has a marginal variance of 1 and an autocorrelation of 0.4. 
		We set $\bm{\beta}=(1,-0.9,0.7,0,0,\dots,0)^T$. 
		The correlated binary responses are generated using R package \texttt{SimCorMultRes}, with  an  exchangeable within-cluster correlation parameter $\rho=0.5\ {\rm or}\ 0.8$.} 
\end{example}

Correlated binary outcomes contain less information than continuous data, making it more challenging to  identify important variables accurately and obtain precise estimates.  The model selection and estimation results for Example \ref{ex2} are presented in Table \ref{binary1}. Notably, the proposed PQL$_{\rm WCR}$ method successfully identifies all significant features and maintains sparsity across all scenarios. By contrast, under an independent correlation structure, PGEE struggles to identify the true model, with a CR value falling below 60\%. While it achieves asymptotic consistency under the other two intracluster correlation structures, this comes at the cost of a high number of FPs. Additionally, the TPs of the naive lasso exhibit instability as the dimensionality $p_n$ increases to 500. With the presence of ICS, the MSE of PQL$_{\rm WCR}$ experiences slight adverse effects but maintains the lowest among the three methods. For $p_n=50$, the PGEE produces highly skewed estimates when observations are assumed to be independent. Even when the working correlation is correctly specified,  the MSE of the PGEE is three times higher than that of PQL$_{\rm WCR}$. Meanwhile, the naive lasso shows significant bias and a large MSE at $p_n=500$.

\begin{table}
	\centering
	\caption{Model selection results for Example \ref{ex2} with ICS.}
	\label{binary1}
	\renewcommand\arraystretch{0.8}
		\vspace{0.2cm}
	\begin{tabular}{ccccccc}
		\toprule 
		$p_n$&$\rho$	&Approach & TP &FP & CR &MSE \\
		\midrule 
		
		50&$0.5$&PQL$_{\rm WCR}$ & 3.00(0.00) & 0.26(0.50) & 1.00(0.00)  & 0.040(0.052)  \\
		~&~&PGEE.indep& 1.77(1.48) & 11.81(10.18)  & 0.59(0.49) & 1.043(1.055) \\
		~&~&PGEE.exch& 3.00(0.00) & 23.57(3.72)  & 1.00(0.00) & 0.136(0.060)  \\
		~&~&PGEE.ar1& 2.85(0.66) & 20.43(5.54) & 0.95(0.22) & 0.256(0.476)  \\
		~&~&naive lasso& 3.00(0.00) & 0.60(0.79)  & 1.00(0.00) & 0.460(0.161)  \\
		
		~&$0.8$&PQL$_{\rm WCR}$ & 3.00(0.00) & 0.23(0.45) & 1.00(0.00)  & 0.048(0.051) \\
		~&~&PGEE.indep& 1.65(1.50) & 11.10(10.44) & 0.55(0.50) & 1.142(1.056)  \\
		~&~&PGEE.exch&3.00(0.00) & 25.46(3.91) & 1.00(0.00) & 0.114(0.049)  \\
		~&~&PGEE.ar1&2.97(0.30) & 23.43(4.46) & 0.99(0.10) & 0.156(0.225)  \\
		~&~&naive lasso& 3.00(0.00) & 0.51(0.69)  & 1.00(0.00) & 0.472(0.169)  \\
		\hline
		500&$0.5$&PQL$_{\rm WCR}$ &3.00(0.00) & 0.03(0.17) & 1.00(0.00)  & 0.101(0.075) \\
		~&~&naive lasso& 2.99(0.10) & 0.44(0.74)  & 0.99(0.10) & 0.833(0.211)  \\
		~&$0.8$&PCRL &3.00(0.00) & 0.03(0.17) & 1.00(0.00)  & 0.115(0.096) \\
		
		~&~&naive lasso& 2.98(0.20) & 0.50(0.81)  & 0.99(0.10) & 0.835(0.217)  \\
		\bottomrule 
		
	\end{tabular}
\end{table}

In Examples \ref{ex3} and \ref{ex4}, we do not introduce ICS so as to investigate the robustness of the proposed PQL$_{\rm WCR}$ method. 

\begin{example}\label{ex3}
	{\rm We consider the linear model as
		\begin{equation*}
			Y_{ij}=\bX^T_{ij}\bm{\beta}+\epsilon_{ij},    
		\end{equation*}
		for correlated normal responses, where $i=1,\dots,200$, $j=1,\dots,M_i$. The values of $M_i$ are generated following the same procedure  as Example \ref{ex1}. The coefficient vector is specified as $\bm{\beta}=(2,-1,1.5,-2,0,0,\dots,0)^T$, representing a homogeneous effect. 		
		The covariates are generated according to the method described in Example \ref{ex1}. 		
		The random errors, $(\epsilon_{i1},\dots,\epsilon_{iM_i})^T$, follow a multivariate normal distribution with a marginal mean of 0, marginal variance of 1, and an exchangeable correlation matrix characterized by a parameter $\rho=0.5\ {\rm or}\ 0.8$.}
\end{example}

Table \ref{continuous2} presents the model selection and estimation results for continuous correlated responses in Example \ref{ex3}. Overall, the proposed PQL$_{\rm WCR}$ method successfully identified all important covariates while achieving the smallest model size among all competing methods. Under the data generation mechanism specified in Example \ref{ex3}, the PGEE indeed provides the optimal estimation, particularly under the assumption of equal correlated. However, given convergence issues in its iterative algorithm, PGEE exhibits several unstable FPs, particularly as the intra-cluster correlation increases. By contrast, the MSEs of  PQL$_{\rm WCR}$ are  comparable with those of PGEE. The naive lasso method selects moderately sized models but produces biased estimates, with the bias amplifying as $p_n=500$. Despite the high dimensionality of covariates exceeding the number of clusters, the proposed PQL$_{\rm WCR}$ method demonstrates robust performance, maintaining stability and effectiveness in model selection and estimation

\begin{table}
	\centering
	\caption{Model selection results for Example \ref{ex3} without ICS.}
	\label{continuous2}
	\renewcommand\arraystretch{0.8}
		\vspace{0.2cm}
	\begin{tabular}{ccccccc}
		\toprule 
		$p_n$&$\rho$	&Approach & TP &FP & CR &MSE \\
		\midrule 
		
		50&$0.5$&PQL$_{\rm WCR}$ & 4.00(0.00) & 0.60(0.79) & 1.00(0.00)  & 0.010(0.010)  \\
		~&~&PGEE.indep& 4.00(0.00) & 2.37(3.56)  & 1.00(0.00) & 0.011(0.010) \\
		~&~&PGEE.exch& 4.00(0.00) & 4.90(5.50)  & 1.00(0.00) & 0.008(0.008)  \\
		~&~&PGEE.ar1& 4.00(0.00) & 4.65(4.79) & 1.00(0.00) & 0.008(0.006)  \\
		~&~&naive lasso& 4.00(0.00) & 2.49(2.11)  & 1.00(0.00) & 0.097(0.042)  \\
		~&$0.8$&PQL$_{\rm WCR}$ & 4.00(0.00) & 0.51(0.72) & 1.00(0.00)  & 0.010(0.009) \\
		~&~&PGEE.indep& 4.00(0.00) & 2.33(3.21) & 1.00(0.00) & 0.009(0.008)  \\
		~&~&PGEE.exch&4.00(0.00) & 9.91(12.92) & 1.00(0.00) & 0.009(0.017)  \\
		~&~&PGEE.ar1&4.00(0.00) & 12.69(11.75) &  1.00(0.00) & 0.008(0.009)  \\
		~&~&naive lasso& 4.00(0.00) & 2.18(1.58)  & 1.00(0.00) & 0.095(0.036)  \\
		\hline
		500&$0.5$&PQL$_{\rm WCR}$ & 4.00(0.00) & 0.19(0.46) & 1.00(0.00)  & 0.009(0.006) \\
		~&~&naive lasso& 4.00(0.00) & 3.09(2.04)  & 1.00(0.00) & 0.207(0.062)  \\
		~&$0.8$&PQL$_{\rm WCR}$&4.00(0.00) & 0.25(0.50) & 1.00(0.00)  & 0.009(0.006) \\
		~&~&naive lasso& 4.00(0.00) & 3.53(2.43)  & 1.00(0.00) & 0.204(0.064)  \\
		\bottomrule 
		
	\end{tabular}
\end{table}

\begin{example}\label{ex4}
	{\rm We model the correlated binary responses with the marginal mean defined as
		\begin{equation*}
			E(Y_{ij}\mid \bX_{ij},M_i)=\exp(\bX^T_{ij}\bm{\beta})/\{1+\exp(\bX^T_{ij}\bm{\beta})\},
		\end{equation*} 
		where $i=1,\dots,200$ and $j=1,\dots,M_i$. The random cluster sizes and covariates are generated following the procedure described in Example  \ref{ex2}.  The coefficient vector is set as $\bm{\beta}=(1,-0.9,0.7,0,0,\dots,0)^T$. 
		Observations within each cluster exhibit an exchangeable correlation structure with parameter $\rho=0.5\ {\rm or}\ 0.8$.}
\end{example}

Table \ref{binary2} summarizes the model selection results and estimation accuracy of the competing methods for Example \ref{ex4}. When the correlated binary responses are independent of cluster sizes, the proposed PQL$_{\rm WCR}$ method consistently outperforms the alternatives in signal identification and parameter estimation.  The PGEE exhibits instability in terms of TPs and FPs across different working correlation matrices. Nevertheless, it shows an improvement in identifying important features compared to Example \ref{ex2}. While the naive lasso method achieves sparsity, it performs worse than our approach due to biased parameter estimation.  As the number of covariates increases to $p_n=500$, the proposed PQL$_{\rm WCR}$ method continues to excel,  selecting an oracle model with the smallest MSE.

\begin{table}
	\centering
	\caption{Model selection results for Example \ref{ex4} without ICS.}
	\label{binary2}
	\renewcommand\arraystretch{0.8}
		\vspace{0.2cm}
	\begin{tabular}{ccccccc}
		\toprule 
		$p_n$&$\rho$	&Approach & TP &FP & CR &MSE \\
		\midrule 
		
		50&$0.5$&PQL$_{\rm WCR}$& 3.00(0.00) & 0.34(0.57) & 1.00(0.00)  & 0.049(0.057)  \\
		~&~&PGEE.indep& 2.10(1.38) & 14.09(9.70)  & 0.70(0.46) & 0.813(0.980) \\
		~&~&PGEE.exch& 2.97(0.30) & 20.57(3.66)  & 0.99(0.10) & 0.173(0.225)  \\
		~&~&PGEE.ar1& 2.88(0.59) & 20.19(5.28) & 0.96(0.20) & 0.249(0.426)  \\
		~&~&naive lasso& 3.00(0.00) & 0.63(0.75)  & 1.00(0.00) & 0.427(0.174)  \\
		~&$0.8$&PQL$_{\rm WCR}$& 3.00(0.00) & 0.25(0.46) & 1.00(0.00)  & 0.043(0.052) \\
		~&~&PGEE.indep& 1.95(1.44) & 12.69(9.67) & 0.65(0.48) & 0.926(1.015)  \\
		~&~&PGEE.exch&3.00(0.00) & 22.94(3.52) & 1.00(0.00) & 0.111(0.048)  \\
		~&~&PGEE.ar1&3.00(0.00) & 22.87(3.75) & 1.00(0.00) & 0.138(0.062)  \\
		~&~&naive lasso& 3.00(0.00) & 0.50(0.77)  & 1.00(0.00) & 0.441(0.167)  \\
		\hline
		500&$0.5$&PQL$_{\rm WCR}$ & 3.00(0.00) & 0.02(0.14) & 1.00(0.00)  & 0.086(0.066) \\
		~&~&naive lasso& 3.00(0.00) & 0.50(0.80)  & 1.00(0.00) & 0.727(0.202)  \\
		~&$0.8$&PQL$_{\rm WCR}$ &3.00(0.00) & 0.06(0.28) & 1.00(0.00)  & 0.091(0.065) \\
		
		~&~&naive lasso& 3.00(0.00) & 0.52(0.85)  & 1.00(0.00) & 0.743(0.212)  \\
		\bottomrule 
		
	\end{tabular}
\end{table}

\section{Yeast Cell-Cycle Gene Expression Data Analysis}\label{real}
In this section, we evaluate the performance of our proposed PQL$_{\rm WCR}$ method using a subset of the yeast cell-cycle gene expression data collected by \cite{Spellman1998}. This dataset includes 292 cell-cycle-regulated genes, with expression levels recorded across two cell-cycle periods. Specifically, repeated measurements were taken on these 292 genes at 7-minute intervals over 119 minutes, resulting in a total of 18 time points. 

The cell-cycle process is typically  divided  into distinct stages: M/G1, G1, S, G2, and M. The M stage (mitosis) involves  nuclear events, such as chromosome separation, and cytoplasmic events, such as cytokinesis and cell division. The G1 stage (GAP 1) precedes DNA synthesis, while the S stage (synthesis) is characterized by DNA replication. The G2 stage (GAP 2) follows synthesis and prepares the cell for mitosis. Transcription factors (TFs) play a crucial role in regulating the transcription of a subset of yeast cell-cycle-regulated genes. \cite{Wang2007} utilized the ChIP data from \cite{Lee2002} to  estimate binding probabilities for these 292 genes, covering a total of 96 TFs.

To identify the TFs potentially involved in the yeast cell-cycle, we consider the following model:
\begin{equation}
	y_{ij}=\sum_{d=1}^{96}\beta_{j,d}x_{ij,d}+\sum_{l_1,l_2=1,\dots,96,l_1<l_2}^{}\gamma_{j,l_1l_2}x_{ij,l_1}x_{ij,l_2}+\epsilon_{ij},\label{yeastmodel}
\end{equation}
where $y_{ij}$ represents the log-transformed gene expression level of gene $i$ measured at time point $j$, for $i=1,\dots, 292$ and $j=1,\dots,18$. The variables $x_{ij,d}$ denotes the matching score of the binding probability of the $d$th TF for gene $i$ at time point $j$. Previous studies have highlighted the importance of interaction effects among certain TFs, such as HIR1:HIR2 \citep{Loy1999}, SWI4:SWI6 \citep{Koch1993}, FKH2:NDD1 \citep{Kumar2000}, SWI5:SMP1 \citep{Banerjee2003}, MCM1:FKH2 \citep{Spector1997}, MBP1:SWI6 \citep{Koch1993} and FKH1:FKH2 \citep{Spector1997,Koranda2000}. Given this evidence, model (\ref{yeastmodel}) is well-justified, resulting in a total of 4,656 covariates to be analyzed.

As discussed in Section \ref{simu}, we compare the selection performance of the proposed PQL$_{\rm WCR}$ method with those of the PGEEs and naive lasso. 
Table \ref{yeast1} presents the identification of 21 TFs that are known and experimentally verified to be related to the cell cycle (referred to as  ``true'' TFs)\citep{Wang2007}. 
Overall, our  PQL$_{\rm WCR}$ method selects the smallest number of TFs while identifying 17 ``true'' TFs, the highest among all competing methods. By contrast, the penalized GEE demonstrates varying performance depending on the assumed within-cluster correlation structures.  When observations are modeled as independent or equally correlated, the PGEE identifies 11  ``true'' TFs out of 60 and 61 selected TFs, respectively.  Under an autoregressive correlation structure, the penalized GEE improves to identify 15  ``true'' TFs among 60 selected TFs. These results highlight the strong dependence of penalized GEE's performance on the choice of working correlation structure. Meanwhile, the naive lasso identifies 12 ``true'' TFs from a comparable total number of selected TFs.

Next, we examine in detail the identification of specific ``true'' TFs. All competing methods successfully select ACE2, CBF1, FKH1, MBP1, NDD1, REB1, STB1, SWI4, and SWI5.
\begin{itemize}
	\item[$\bullet$] BAS1: Known to regulate the synthesis of histidine, purines, and pyrimidines \citep{Rao2011}, BAS1 also plays a role in preparing yeast cells for division. However, the penalized GEE fails to identify it as important.
	\item[$\bullet$] MCM1: This TF primarily exerts its regulatory effects during the M/G1, S, and early G2 stages, indicating its critical role in the yeast cell cycle. Identified as one of the three major TF types by \cite{Spellman1998}, MCM1's vital impact is overlooked by the penalized GEE.
	\item[$\bullet$] MET31: Required in interaction with the activator MET4 to bind DNA, MET31 regulates sulfur metabolism \citep{Carrillo2012}, a process linked to the initiation of cell division \citep{Blank2009}. Notably, MET31 is exclusively identified by the PQL$_{\rm WCR}$ method.  
\end{itemize}
The naive lasso, despite achieving sparsity, fails to select several significant TFs with substantial biological roles. For instance:
\begin{itemize}	  	
	\item[$\bullet$] FKH2: plays a dominant role in regulating genes associated with nuclear migration and spindle formation \citep{Pic2000}. 
	\item[$\bullet$] SWI6: coordinates gene expression at the G1-S boundary of the yeast cell cycle, as noted by \cite{Foord1999}.
\end{itemize}

These results underscore the superior performance of our proposed PQL$_{\rm WCR}$ method in reliably identifying critical TFs relevant to the yeast cell cycle.
However, there are a few TFs that were not successfully detected by  PQL$_{\rm WCR}$. Notably, GCN4 was not identified as significant by any of the methods considered. The naive lasso method was the only approach to select GCR1, while SKN7 and STE12 were recognized as cell cycle regulators only by the PGEE. To understand these discrepancies further, we turn to relevant biological research and offer plausible explanations.

First, heterogeneous regulatory effects of TFs during the yeast cell cycle likely contribute to these observations. Different TFs exhibit varying levels of regulatory involvement at different stages of the cycle. Moreover, some TFs regulate the cell cycle in a periodic manner, with their effects confined to specific intervals. Given the PQL$_{\rm WCR}$ method samples observations at arbitrary time points, it tends to select TFs with consistent regulatory effects throughout the two observed periods. For instance, STE12 is known to direct the cyclic expression of certain genes during the early G1 stage, as described by \cite{Oehlen1996}. This stage represents a relatively narrow time interval, resulting in a low sampling probability and an estimated coefficient close to zero. However, if observations were sampled exclusively during the G1 stage, STE12 could be successfully identified.

Additionally, the intricate interactions between TFs may influence the identification results. For example, SKN7 exhibits several genetic interactions with MBP1 during the G1-S transition, including mutual inhibition \citep{Bouquin1999}. Despite this interaction, our PQL$_{\rm WCR}$ method identified MBP1 as significant, potentially overshadowing SKN7.

In addition to the 21 verified TFs, the PQL$_{\rm WCR}$ method also identified additional regulatory TFs supported by biological evidence of their roles in the yeast cell cycle. For instance, HIR1 is known to participate in the cell cycle?dependent repression of histone gene transcription \citep{Osley1987}. Similarly, \cite{Moser2013} found that PHD1 contributes to centriole duplication and centrosome maturation, playing a role in the regulation of cell cycle progression.

\begin{table}
	\centering
	\caption{Results for identifying the 21 "true" TFs. "1" means that the TF is selected, while "0" means the opposite.}
	\renewcommand\arraystretch{0.8}
	\label{yeast1}
		\vspace{0.2cm}
	\begin{tabular}{cccccc}
		\toprule  TF & PQL$_{\rm WCR}$ & PGEE.indep  &PGEE.exch & PGEE.ar1 & naive lasso\\
		\midrule
		ABF1 & 1  &1 &1&1&0\\
		ACE2 & 1  &0 &0&1&1\\
		BAS1 & 1  &0 &0&0&1\\	
		CBF1 & 1 &1 &1&1&1\\
		FKH1 & 1 &1 &1&1&1\\
		FKH2 & 1 &1 &1&1&0\\
		GCN4 & 0 &0 &0&0&0\\
		GCR1 & 0 &0 &0&0&1 \\
		GCR2 & 1 &0 &0&1&0 \\
		LEU3 & 1 &1 &1&0&0 \\
		MBP1 & 1 &0 &0&1&1 \\
		MCM1 & 1 &0 &0&0&1\\
		MET31 & 1 &0 &0&0&0\\
		NDD1 & 1 &1 &1&1&1\\ 
		REB1 & 1  &1 &1&1&1\\		
		SKN7 & 0  &0 &0&1&0\\
		STB1 &  1 &1 &1&1&1\\
		STE12 & 0 &1 &1&1&0 \\
		SWI4 & 1 &0 &0&1&1\\
		SWI5 & 1 &1 &1&1&1\\
		SWI6 & 1 &1 &1&1&0\\								    	
		\hline
		Number of "true" TFs&17&11 &11&15&12\\
		Total number of TFs&52&60 &61&60&60\\
		\bottomrule
	\end{tabular}
	
\end{table}

\section{Discussion}\label{discu}
In this article, we tackle the challenges of variable selection and estimation in high-dimensional longitudinal data with ICS. Leveraging the technique of WCR, we develope a PQL method applied to each resampled dataset. Theoretical analysis establishes the consistency of model selection and estimation for this approach. To integrate the estimators from resampled datasets, we introduce a penalized mean regression technique, yielding an aggregated estimator that enhances true positive discovery while reducing false positives.  This integrated methodology, termed PQL$_{\rm WCR}$, combines the strengths of WCR and PQL. Simulation studies demonstrate that  PQL$_{\rm WCR}$ achieves consistent model selection and robust estimation, even when outcomes are dependent on cluster size. Consequently, PQL$_{\rm WCR}$  provides an efficient and reliable tool for analyzing longitudinal data, requiring only assumptions about the first and second moments of the data.

\section*{Acknowledgements}
This research was supported by NSFC Grant No. 12271238, Guangdong NSF Fund 2023A1515010025, and Shenzhen Sci-Tech Fund (JCYJ20210324104803010) for Xuejun Jiang.

\section*{Declarations}

\textbf{Conflict of interest} The authors declare that they have no conflict of interest.

\newpage

\end{document}